\documentclass[a4paper,11pt]{article}
\pdfoutput=1 

\usepackage{jinstpub} 

\usepackage{lineno}

\title{\boldmath 
Detailed analysis of chemical corrosion of ultra-thin 
 wires used in drift chamber detectors.}

\author[a]{A.M.Baldini,}
\author[d]{G.Cavoto,} 
\author[b,1]{F.Cei,\note{Corresponding author.}}  
\author[a]{M.Chiappini,} 
\author[d]{G.Chiarello,} 
\author[f]{C.Chiri,} 
\author[g]{G.Cocciolo,} 
\author[f]{A.Corvaglia,} 
\author[g]{F.Cuna,} 
\author[b]{M.Francesconi,}
\author[a]{L.Galli,} 
\author[f]{F.Grancagnolo,} 
\author[a]{M.Grassi,}
\author[e]{R.Ishak,} 
\author[d]{M.Meucci,} 
\author[b]{D.Nicol\'{o},} 
\author[g]{M.Panareo,} 
\author[b]{A.Papa,}
\author[g]{A.Pepino,}
\author[a]{F.Raffaelli,}
\author[c]{F.Renga,} 
\author[d]{E.Ripiccini,}
\author[a]{G.Signorelli,}
\author[f]{G.F.Tassielli,}
\author[e]{R.Valentini,}
\author[c]{C.Voena.}

\affiliation[a]{INFN Sezione di Pisa \\Largo Bruno Pontecorvo, 3, 56127, Pisa, Italy}
\affiliation[b]{University of Pisa, Department of Physics and INFN Sezione di Pisa \\Largo Bruno Pontecorvo, 3, 56127, Pisa, Italy}
\affiliation[c]{INFN Sezione di Roma, \\Piazzale A. Moro, 2, 00185, Rome, Italy}
\affiliation[d]{University of Rome, La Sapienza Department of Physics and INFN Sezione di Roma, \\Piazzale A. Moro, 2, 00185, Rome, Italy}
\affiliation[e]{University of Pisa, Department of Mechanical Engineering, \\Via Diotisalvi, 2, 56122, Pisa, Italy}
\affiliation[f]{INFN Sezione di Lecce, \\Via per Arnesano, 73100, Lecce, Italy}
\affiliation[g]{University of Salento, Department of Mathematics and Physics and INFN Sezione di Lecce, \\Via per Arnesano, 73100, Lecce, Italy}

\emailAdd{fabrizio.cei@unipi.it}

\abstract{Ultra-thin metallic anode and cathode wires are frequently employed in low-mass gaseous detectors for precision experiments, where the amount of material 
crossed by charged particles must be minimised. We present here the results of an analysis of the mechanical stress and chemical corrosion effects observed in $40$ and $50~{\rm{\mu m}}$ diameter silver plated aluminum wires mounted within the volume of the MEG\,II drift chamber, which caused the breakage of about one hundred wires (over a total of $\approx 12000$). This analysis is based on the careful inspection of the broken wires by means of optical and electronic microscopes and on a detailed recording of all breaking incidents. We present a simple empirical model which relates the number of broken wires to their exposure time to atmospheric relative humidity and to their mechanical tension, which is necessary for mechanical stability in the presence of electrostatic fields of several kV/cm. Finally we discuss how wire breakages can be avoided or at least strongly reduced by operating in controlled atmosphere during the mounting stages of the wires within the drift chamber and by choosing a $25\,\%$ thicker wire diameter, which has very small effects on the detector resolution and efficiency and can be obtained by using a safer fabrication technique.} 


\begin{document}
\maketitle
\flushbottom

\section{Introduction}
\label{sec:intro}
\begin{figure}[htbp]
\centering 
\includegraphics[width=.6\textwidth,trim=0 0 0 0,clip]{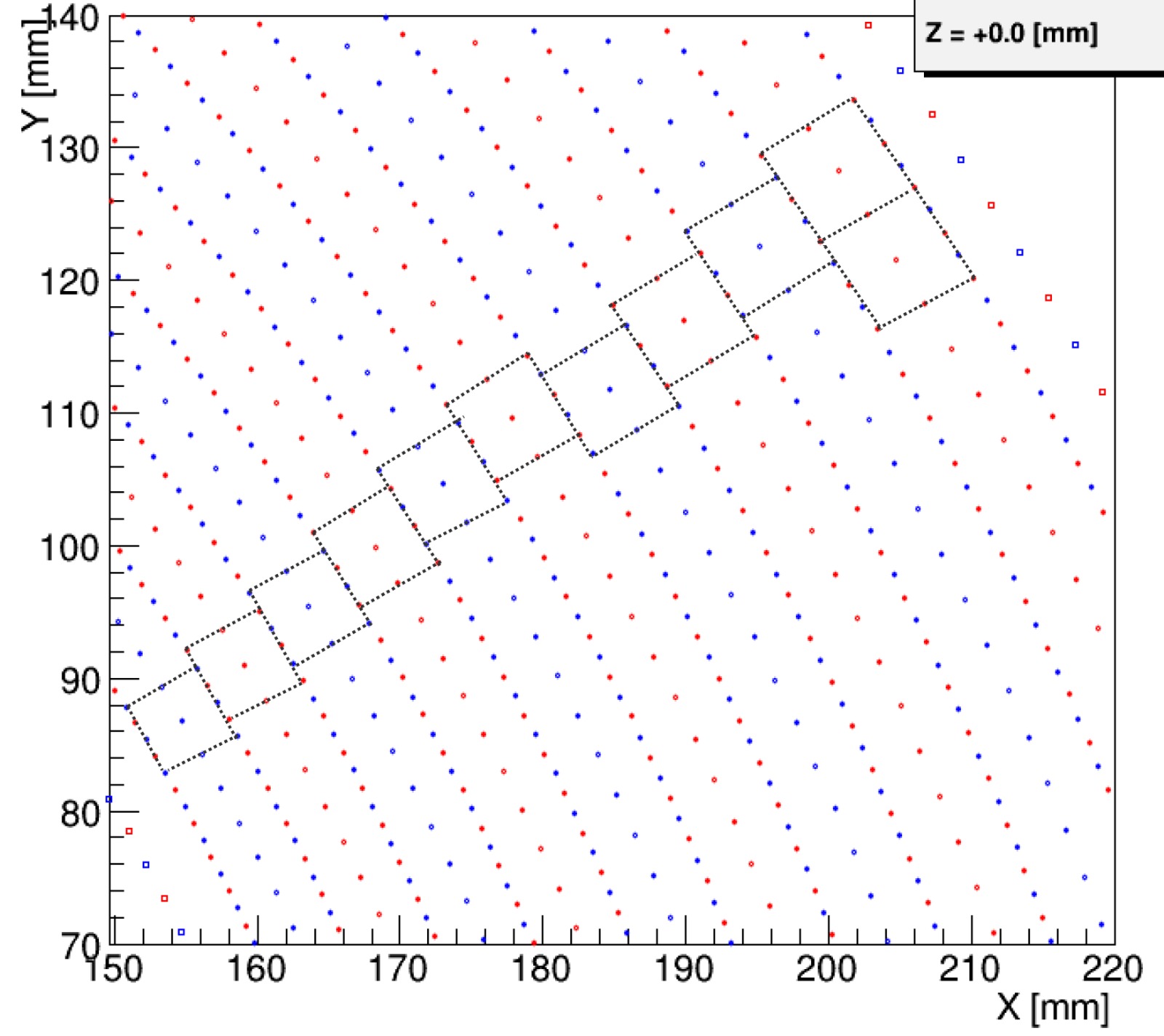}
\caption{\label{fig:DCHLayout} Schematic layout of the drift cell and wire configuration at the chamber centre.}
\end{figure}

Low mass gaseous detectors are frequently used in particle physics precision experiments, where the amount of material crossed by charged particles must be minimised. On the other hand, since accurate measurements of particle momenta are mandatory, these detectors are usually equipped with dense arrays of anode and cathode wires, which must be as thin as possible to avoid a resolution worsening due to multiple scattering and energy loss. The MEG\,II experiment~\cite{MEGUpgrade} is an upgraded version of the MEG experiment~\cite{MEGDet}. MEG established the most stringent limit on the Lepton Flavour Violating process $\mu^{+} \rightarrow {\rm e^{+}} \gamma$~\cite{MEGResult} and a very stringent limit on the process $\mu \rightarrow {\rm e^{+}} \gamma \gamma$~\cite{MEG2Gamma} and performed high statistics studies of muon properties~\cite{MEGRD, MEGPol}. In the MEG\,II detector the measurement of the charged particle momenta is performed by a very low mass drift chamber~\cite{MEGIINIMA, MEGIIJINST, MEGIIJINST2}, filled with a 90:10 ${\rm He-iC_{4}H_{10}}$ gas mixture and equipped with $1728$ gold plated Tungsten anode wires, $7680$ $\phi = 40~{\rm{\mu m}}$ and $2496$ $\phi = 50~{\rm{\mu m}}$ silver plated aluminum wires\footnote{$\phi$ is the conventional symbol to indicate the diameter of a cylindrical piece.}. The drift chamber volume is immersed in a longitudinally varying magnetic field, with its maximum value ($1.26~{\rm T}$) at the detector centre and about a factor two lower at the longitudinal limits of the chamber. The magnetic field forces the positrons emitted in muon decays to follow an helicoidal trajectory; the projection of this trajectory in the plane orthogonal to the chamber axis is a group of circular paths, usually more than one, called \lq\lq turns\rq\rq. 
Tungsten anode and cathode wires are mounted in a stereo geometry to allow a three-dimensional particle track reconstruction; aluminum wires are used as cathodes or placed between two consecutive anode wires and close to the mechanical structure of the chamber in order to define the electric field at the boundaries of the gaseous volume (guard wires). The wire length is $\approx 1.93~{\rm m}$. The drift cells are approximately squares with a side of $\approx 7~{\rm mm}$. A layout of the drift cell and wire configuration at the chamber centre is shown in figure~\ref{fig:DCHLayout}.
The total amount of material crossed on average by a $52.83~{\rm MeV}$ positron (the energy of the positron in the possible $\mu \rightarrow {\rm e^{+}} \gamma$ process and the maximum energy of the positron in the usual muon decay $\mu^{+} \rightarrow {\rm e^{+}} \bar{\nu}_{\rm \mu} \nu_{\rm e}$) is $1.6 \times 10^{-3}~X_{0}$ per track turn and the expected momentum and angular resolutions at $52.8~{\rm MeV}$ are respectively $\sim 100~{\rm keV}$ and $\sim 6.5~{\rm mrad}$. 

The MEG\,II drift chamber mounting procedures~\cite{MEGIIJINST2} started in $2016$. The mechanical structure consists of a pair of 
$3~{\rm cm}$ thick circular aluminum end-plates kept in position at an adjustable distance by a central steel shaft which could be removed after the end of the wiring phase when a carbon fiber cover was externally screwed to the end-plates keeping them at the correct distance. Wiring consisted of soldering both ends of groups of thirty-two wires to metallic pads deposited on printed circuit boards (PCBs) which were subsequently fixed, in concentric layers, to the end-plates, duly machined for keeping PCBs in position. 
Soldering wires on PCBs was accomplished  by means of computer controlled positioning devices, solder paste feeder and laser head. PCBs and wires mounting in the drift chamber proceeded from inner (near the steel shaft) layers to outer layers.
Before being mounted in the chamber, PCBs with soldered wires were stored on two meters long aluminum trays in a nitrogen atmosphere. Wires soldering to PCBs is essential in this construction strategy; this is why silver plated aluminum wires were used. 

PCBs mounting in the drift chamber was done in a class $10000$ clean room environment which was temperature ($25^{\circ}~$C) but not relative humidity controlled; wires were exposed to atmospheric relative humidity which could typically range between $45$ and $65\,\%$. Mounting of a full wire layer required $\sim 5$ days. Given the $\sim 1.5~{\rm kV}$ potential difference needed for gas gain purposes between anodes and cathodes, during normal operation wires are stretched $5.2~{\rm mm}$ beyond their rest length to avoid short-circuits when the potential difference is switched on. This stretch value was obtained by Garfield~\cite{Garfield} based simulations and experimentally verified. During mounting, wires in the chamber were usually left
stretched\footnote{We will omit from here on that the stretch is relative to the rest length.} at 2 mm. In summer 2017, when  wiring was $70\,\%$ completed, we observed the breaking of the first few wires, mainly in the inner layers.
The chamber was completed in Spring $2018$ and then moved to the PSI laboratory~\cite{PSI}, where the MEG\,II experiment is located. Several other breakages occurred later on 
when the chamber was stretched up to $6~{\rm mm}$.   
The total number of $\phi = 40~{\rm{\mu m}}$ aluminum broken wires is $97$ while that of $\phi = 50~{\rm{\mu m}}$ wires is ten; no anode tungsten wire ever broke. Since the number of broken cathodes is less than $1\%$ of the total, one can expect the influence on the track reconstruction efficiency to be not so dramatic; this is also the case since, given 
the stereo geometry of wires and
the redundancy of cathodes in a single drift cell, we verified by means of ANSYS Electronic desktop~\cite{ANSYS} simulation that the loss of one $40~\rm{\mu m}$ cathode does not change the cell electric field appreciably. 

It is however necessary to completely remove broken wires fragments from the chamber since they could cause electrostatic discharges or short circuits. Since the average anode-cathode distance is around $3~{\rm mm}$ this task is far from being trivial. It was accomplished by means of a set of hand controlled micro movements (four  rotations and a linear movement) of a thin ($1~{\rm mm}$ diameter) steel hook guided by two cameras with large magnification, with which all the broken wires pieces were carefully recuperated without breaking any further wire.
This paper reports on the detailed inspection and analysis of all the wire breakages. Since a new backup chamber construction has already started, the analysis presented here was fundamental in order to build and operate the new chamber in a much safer way to avoid or at least to minimise wire breakages and their undesirable effects. 

\begin{figure}[htbp]
\centering 
\includegraphics[width=.9\textwidth,trim=30 110 0 0,clip]{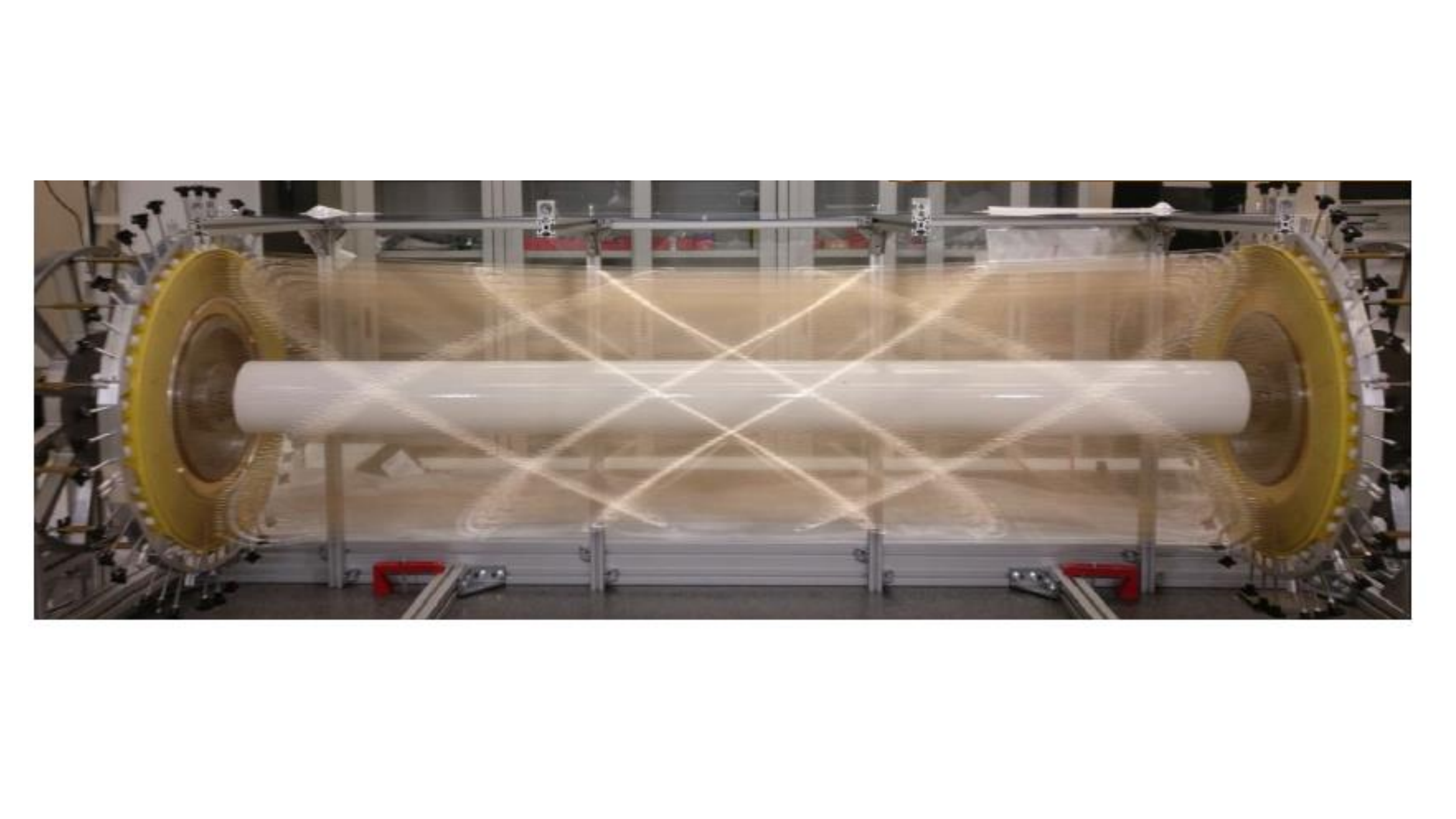}
\caption{\label{fig:MEGIIDCH} The MEG\,II experiment drift chamber.}
\end{figure}

The MEG\,II drift chamber is shown in figure~\ref{fig:MEGIIDCH} and described in detail in~\cite{MEGUpgrade}; further information on wires and gas mixture characteristics, mounting procedures, readout and trigger electronics can be found in~\cite{MEGIINIMA, MEGIIJINST, MEGIIJINST2} and in~\cite{MEGIIDAQ, MEGIITrigger}.
\section{Characteristics of the ultra-thin wires used in the MEG\,II drift chamber.}
\label{sec:properties}
\begin{figure}[htbp]
\centering 
\includegraphics[width=.45\textwidth,trim=0 0 0 0,clip]{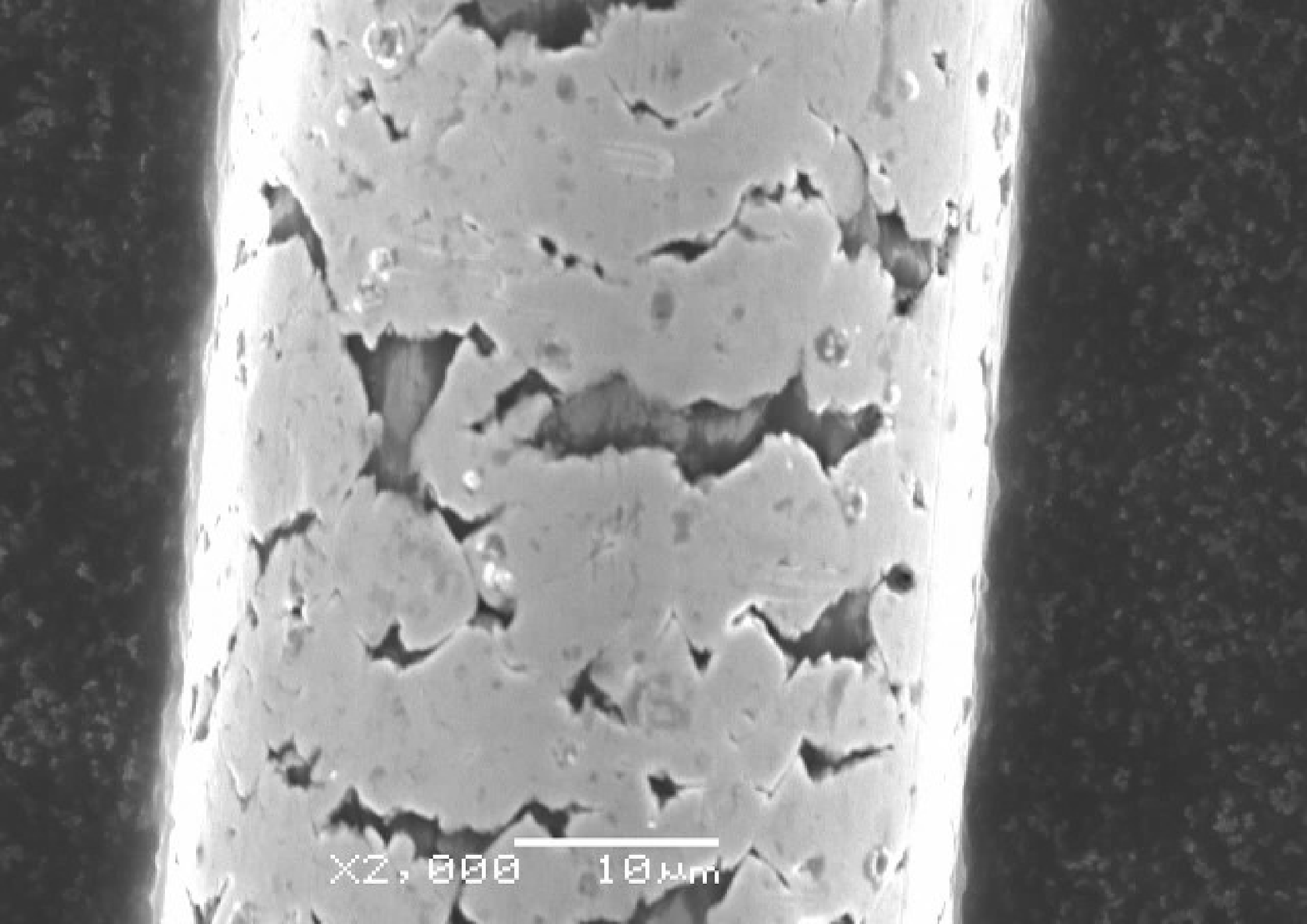}
\qquad
\includegraphics[width=.45\textwidth,trim=0 0 0 0,clip]{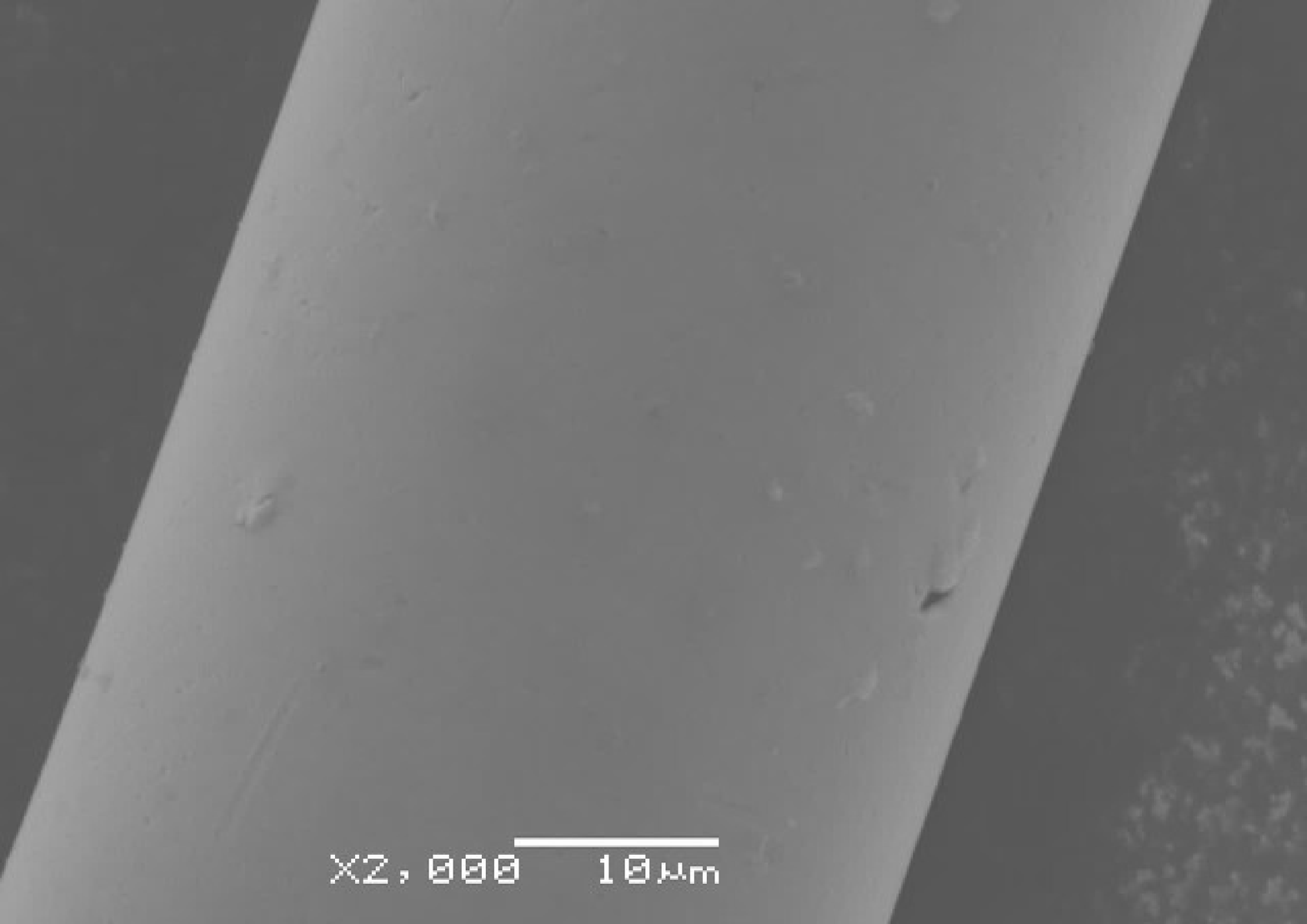}
\caption{\label{fig:cracks} Examples of electronic microscope pictures showing cracks in the $\rm{Ag}$ coating of silver plated aluminum wires. A $\phi = 40~{\rm{\mu m}}$ wire is shown on the left, a $\phi = 50~{\rm{\mu m}}$ one on the right. The $\phi = 40~{\rm{\mu m}}$ wire, which was obtained after a more stressing ultra-finishing procedure, has a much higher density of cracks. Both pictures were taken with a $\times 2000$ magnification, as reported on them.}
\end{figure}

The wires employed in the MEG\,II drift chamber were produced in sets of several spools by the California Fine Wire (CFW)~\cite{CFW} company by using an initial wire drawing procedure which decreases the wire diameter to $\phi = 55~{\rm{\mu m}}$ in several stages. A 
final step, called the ultra-finishing procedure, is used for reaching the design value of $\phi = 40~{\rm{\mu m}}$ or $\phi = 50~{\rm{\mu m}}$. The initial drawing procedure is only effective for wire diameters larger than $50~{\rm{\mu m}}$ therefore ultra-finishing is unavoidable for $\phi = 40~{\rm{\mu m}}$ wires but not for $50~{\rm{\mu m}}$ wires. 
Ag coating is performed by electrochemical deposition before the ultra-finishing stage, therefore this coating is exposed to a mechanical stress which can produce localised cracks.
This effect is particularly important for $\phi = 40~{\rm{\mu m}}$ wires where the diameter reduction during ultra-finishing must be three times larger than that in the 
$50~{\rm{\mu m}}$ case. Examples of cracks on $\rm{Ag}$ coating are shown in figure~\ref{fig:cracks} for a $\phi = 40~{\rm{\mu m}}$ wire on the left and for $\phi = 50~{\rm{\mu m}}$ wire on the right. 
Comparison between these two wires clearly shows that the $\phi = 40~{\rm{\mu m}}$ wire has a much higher density of cracks in its coating than the $\phi = 50~{\rm{\mu m}}$ wire. Moreover the density of these cracks varies considerably depending on the particular wire spool used. We will discuss later the role of cracks in the development of the corrosion process and the advantages of avoiding the ultra-finishing step. 

When a long mechanical piece, such a wire or a rod, is exposed to an external longitudinal tension force, it experiences a stretch, which can be permanent or not. The amount of the tension force is usually given in terms of the ratio between the force itself and the cross sectional area of the piece; this ratio is called \lq\lq stress\rq\rq. If the piece lengthening is proportional to the stress intensity and it shrinks back to its rest length when the tension is no more applied, the force intensity lies within the {\it elastic region} of the piece and the constant ratio between the external force and the stretch is called the \lq\lq elastic constant\rq\rq~of the piece. However, above a certain stress the lengthening value is no longer proportional to the tension intensity and the piece is unable to recover its rest length when the force is no longer applied: this is called the {\it plastic region} of the piece. A further increase of the stress causes the piece to break; the minimum stress which produces the breaking is called the \lq\lq breaking stress\rq\rq~and the piece lengthening where the breaking takes place is called the \lq\lq breaking stretch\rq\rq. 

Given the MEG\,II drift chamber construction method and the small distances between anodes and cathodes, it was important to avoid plastic deformation of wires which could have led not only to an uncertainty in the wires' position, undesirable from the track reconstruction point of view but also to possible delayed electric shorts. The fabrication procedure of the MEG\,II wires was therefore chosen so that their plastic region be as small as possible. As a result, the breaking stretch of the MEG\,II wires is just above the upper limit of their elastic region, i.e. the elastic limit stretch. The elastic limit range for MEG\,II wires is $6.7 - 8.8~{\rm mm}$ and the breaking stretch range is $9.0 - 10.5~{\rm mm}$, so that the 6 mm maximum stretch which was applied to the wires in the chamber corresponds to $\approx 75\,\%$ of the average value of the elastic limit and to $\approx 63\,\%$ of the average value of the breaking stretch.
\begin{figure}[htb]
\centering 
\includegraphics[width=.9\textwidth,trim=0 0 0 0,clip]{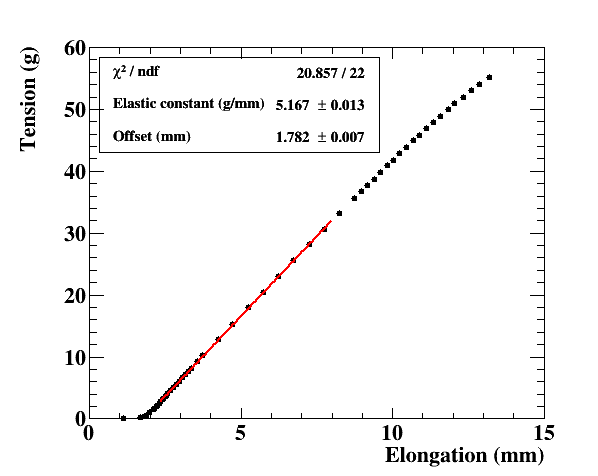}
\caption{\label{fig:Filoca} Tension vs stretch curve measured for a $40~{\rm{\mu} m}$ cathode wire interpolated with a straight line fit. The fitted parameters are the wire elastic constant and the offset in the stretch length (see the text for the meaning of this offset).}
\end{figure} 
We show in figure\,\ref{fig:Filoca} the measured tension versus stretch curve for a $40~{\rm{\mu} m}$ cathode 
where the fitted values are the wire elastic constant and the offset 
in the stretch length. This offset is not zero since the elongation is measured from a fixed position below the wire rest length; the true stretch length with respect to the rest length is therefore the difference between the measured elongation and the offset.    
\section{Inspection of broken wires.}
\label{sec:SEM}
Most fragments of broken wires were studied by using optical microscopes and Scanning Electron Microscopes (SEM) at INFN Pisa, Lecce, at the Department of Mathematics and Physics  of Lecce University, at the Department of Mechanical Engineering of Pisa University and at the Engineering Laboratories of Bergamo University. 
\begin{figure}[htbp]
\centering 
\includegraphics[width=.8\textwidth,trim=0 0 0 0,clip]{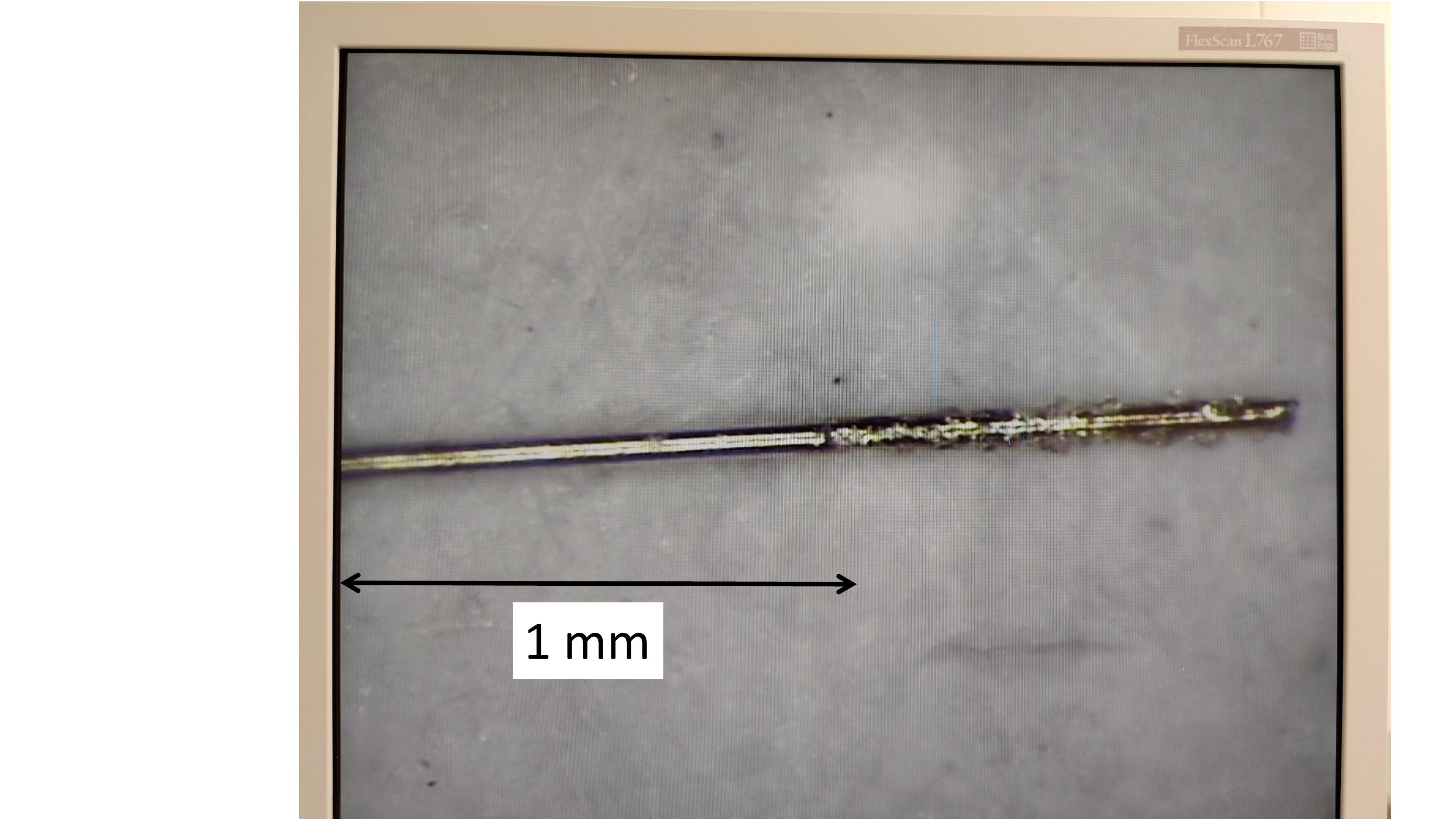}
\caption{\label{fig:Ottico} Optical microscope picture of a $40~{\rm{\mu m}}$ broken wire: magnification around $\times 300$.}
\end{figure} 
Figure~\ref{fig:Ottico} shows a typical example of a broken $40~{\rm{\mu} m}$ wire as seen under the optical microscope. 
We note the very clean cut of the wire at the breaking point, without any plastic deformation; the damage of the silver deposit near the breaking, with the presence of a whitish fluff; the improvement of the conditions of the deposit at distances larger than around $1~{\rm mm}$ from the breaking.

\begin{figure}[htbp]
\centering 
\includegraphics[width=1.1\textwidth,trim=20 150 0 0,clip]{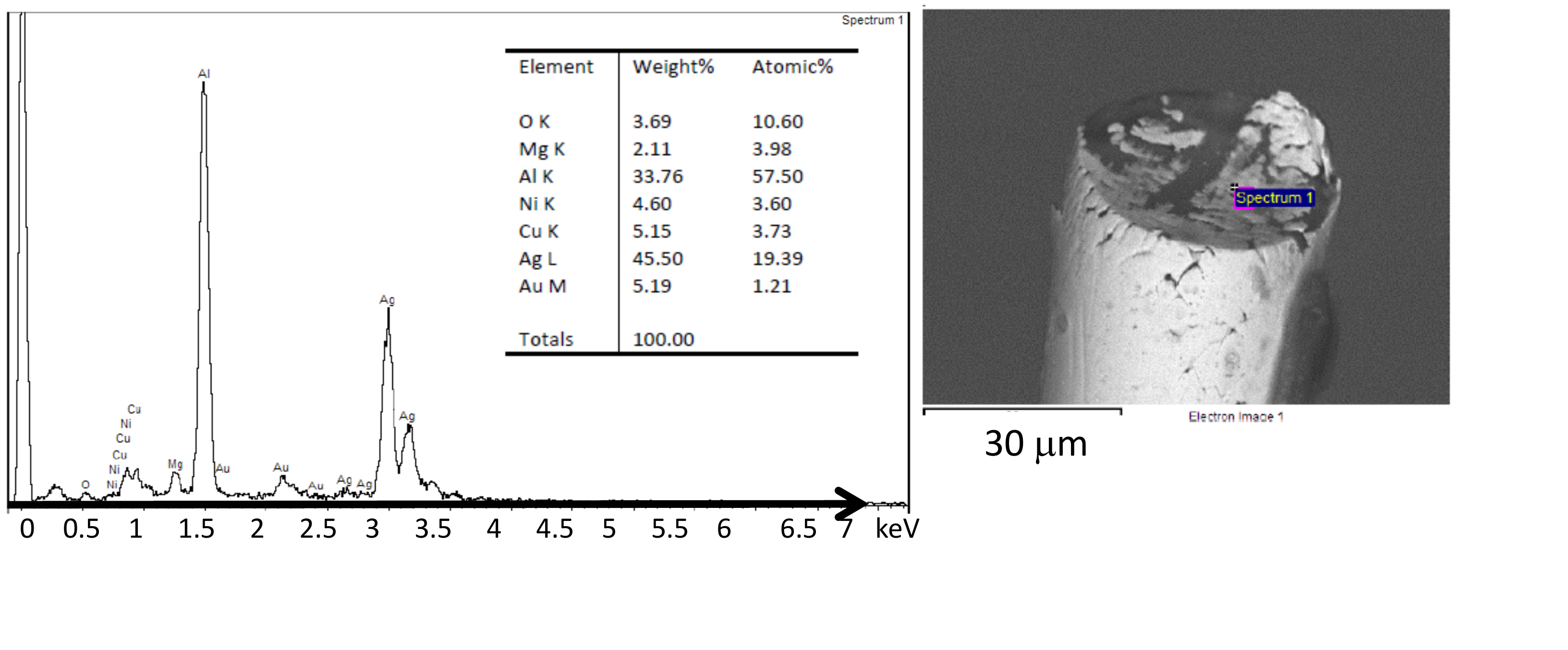}
\includegraphics[width=1.1\textwidth,trim=20 150 0 0,clip]{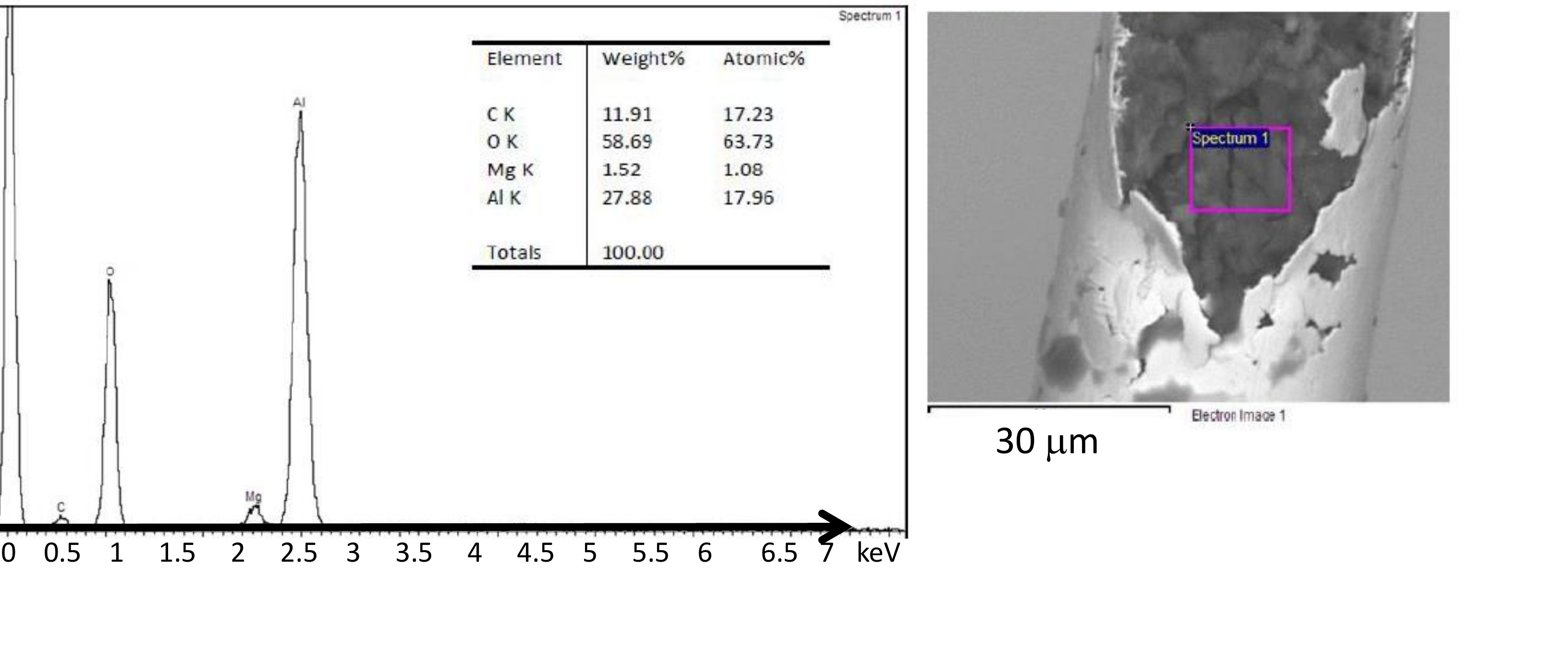}
\caption{\label{fig:phot1-2} Examples of pictures of broken wires taken with a 
$\times 2500$ magnification SEM. The results of the X-ray spectroscopic analysis and the relative concentrations of the materials extracted from the spectral analysis are also reported. The energy scales are in keV.}
\end{figure}

Pictures of broken wires were also taken with the SEM;
in this case chemical composition of the wire surface in points close to the breakages could also be derived by energy dispersive X-ray spectroscopy (EDS). Examples of these broken wires with the corresponding chemical composition are shown in figure~\ref{fig:phot1-2}.
SEM analysis highlights the presence of extensive corrosion phenomena close to the breaking points. Abundant corrosion products in the form of bubbles and filaments can be observed on the surfaces of the broken wires; such products are localised at the interface between silver and aluminum and cause the coating to bloat. Cracks and discontinuities on the silver coating can also be seen, making the aluminum core of the wire clearly visible; these discontinuities were probably dilated by  wire stretch. Chemical composition sometimes shows the presence of elements not strictly related to the aluminum or silver alloys used for wire fabrication; traces of chlorine and other halogen elements are visible in some cases.    

All these observations favour wire breakages induced by the development of corrosion processes in the aluminum alloy, under the silver coating. This kind of corrosion and breaking processes is extensively discussed in technical literature~\cite{corrpaper}. At room  temperature the onset of corrosion processes requires the presence of water or water vapour condensed on the aluminum surface. A humid environment is then the natural scenario in which the corrosion can start and proceed. The minimum relative humidity percentage needed for the condensation depends on several factors; 
it is usually presumed~\cite{BergamoUniv} that this limit is $\sim 60\,\%$, but it can be significantly lowered in localised sections of the wire surfaces with higher porosity. The $60\,\%$ relative humidity is in any case not far from the level which was normally present during the chamber assembly. Local contaminations of hygroscopic materials, like calcium or magnesium salts, whose presence was sporadically observed close to some breaking points, can also trigger water vapour condensation even at lower relative humidity values. A \lq\lq positive feedback\rq\rq~corrosion process takes place when chlorine is present, since it can interact with alkaline elements to form hygroscopic salts. Nevertheless, the water vapour is by itself sufficient to produce corrosion processes on aluminum alloys, since hydrogen ions are easily liberated by water dissociation and formation of aluminum oxide. The cracks in the silver coating are particularly dangerous, since they expose the aluminum alloy to the environmental relative humidity. The galvanic coupling between silver and aluminum and the presence of a large percentage of magnesium in the aluminum alloy $5056$ used for wire fabrication make possible a rapid and severe corrosion attack in localised regions close to the cracks.  

In summary all wire inspections favour the corrosion process at the silver-aluminum interface as the cause of the wire breakages and can provide some possible recipes to mitigate or possibly eliminate these kinds of problems. We will discuss these suggestions in section~\ref{sec:future}. 
\section{An empirical model for corrosion development.}
\label{sec:ana}
A detailed statistical analysis was performed on the broken wires, looking at possible correlations between the number of broken wires, the breaking point position along the wire and in radial direction from the chamber axis, the average exposure time to the humid atmosphere, the broken wire diameter, the wire stretch length when the breakages took place etc. 

As already noted, the total number of broken wires was $107$, $97$ of $\phi = 40~{\rm{\mu m}}$ and $10$ of $\phi = 50~{\rm{\mu m}}$. Taking into account the total number of installed wires of the two types ($7680$ for $40~{\rm{\mu m}}$ and $2496$ for $50~{\rm{\mu m}}$ wires), one can estimate a relative breaking probability $P_{50/40}$ as: 
\begin{equation}
\label{eq:relprob}
P_{50/40} = \frac{10}{97} \times \frac{7680}{2496} = 0.32 \pm 0.11. 
\end{equation}
where the uncertainty is computed by using the Poisson fluctuations of the number of 
$\phi = 40~{\rm{\mu m}}$ and of $\phi = 50~{\rm{\mu m}}$ broken wires.
A simple scaling based on the cross section ratio 
of the two wires would suggest a relative breaking probability of $0.64$: a factor two larger. This clearly indicates that the wire diameter is not enough to explain the different number of breakages occurred for $40$ and $50~{\rm{\mu m}}$ diameter wires. 

In the rest of this section we will consider only the breakages of the $40~{\rm{\mu m}}$ wires because they constitute a homogeneous sample from the production point of view and a much larger statistical sample than the few $50~{\rm{\mu m}}$ wires broken.

We show in table~\ref{tab:nvslambda} the integral number of broken wires as a function of the stretch, i.e. the number $N_{broken} \left( \lambda \le \bar{\lambda} \right)$ of wires which broke for a stretch length lower than or equal to $\bar{\lambda}$ vs $\bar{\lambda}$. Stretches were performed at different times in order to be able to remove only a few wires at a time. Removing a large number of broken wires at the same time is problematic since wires tend 
to tie to one another. Stretch was performed above the final $5.2~{\rm mm}$ value to be sure that no more wires would break when the chamber is closed and in operation.
\begin{table}[htbp]
\centering
\caption{\label{tab:nvslambda} Integral spectrum of the number of broken wires as a function of the stretch $\bar{\lambda}$.}
\smallskip
\begin{tabular}{|c|c|}
\hline
$\bar{\lambda}~\left( mm \right)$ & $N_{broken} \left( \lambda \le \bar{\lambda} \right)$ \\
\hline
2.0 & 6   \\
3.0 & 12 \\
3.3 & 15 \\
3.8 & 25 \\
4.6 & 41 \\
5.0 & 56 \\
5.5 & 63 \\
6.0 & 97 \\
\hline
\end{tabular}
\end{table}

Table~\ref{tab:nvslambda} shows that the number of broken wires has a strong dependence on the stretch;
this dependence can be phenomenologically approximated
with a power law (with power exponent $3.5 - 4$) or with an error function. In both cases one can extrapolate that the total number of wires which would break by a further stretch up to the elastic limit is $\sim 250$, about $3\,\%$ of all the $40~{\rm{\mu m}}$ diameter wires. This can be considered as the total amount of wires in which corrosion started during the assembly phase.

\begin{figure}[htb]
\centering 
\includegraphics[width=.6\textwidth,trim=0 0 0 0,clip]{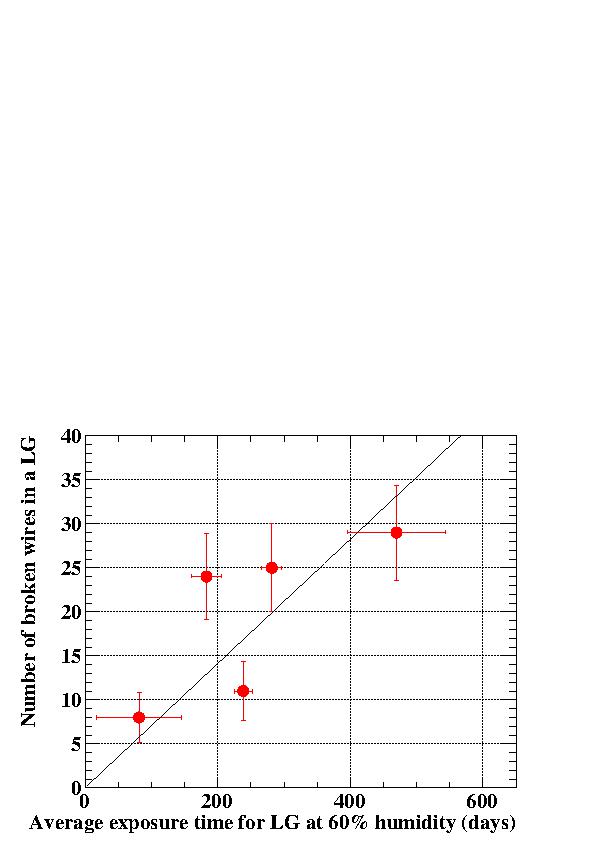}
\caption{\label{fig:humid} Number of broken wires in groups of two consecutive layers (LG: Layer Group) as a function of average time in $60\,\%$ equivalent relative humidity.}
\end{figure}

An accurate evaluation of the exposure time of broken wires to relative humidity is not easy, due to the very different conditions during the assembly phase which lasted around two years. Nevertheless, reasonable estimates of the average exposure time to $60\,\%$ equivalent relative humidity indicate that the number of broken wires is roughly proportional to this exposure time
as shown in figure\,\ref{fig:humid} where broken wires are grouped for two consecutive layers (LG: Layer Group) to reduce the experimental uncertainties. This dependence represents a strong confirmation of the validity 
of the corrosion hypothesis.

We developed a phenomenological model of the 
wire breakages which combines timing and stretch information. The purpose of this model is to provide a quantitative tool to predict the expected number of broken wires in the present and in a new drift chamber. In this section we describe the characteristics of this model and its performance and in the following one we will apply it to the strategy planned for the preparation of the new chamber.    

An elastic wire can be thought of as a huge number of very small springs, uniformly distributed across its cross section. Since the elastic constant of a group of springs mounted in parallel is the sum of their elastic constants, we can think of the elastic constant of the wire as due to the sum of the elastic constants of these \lq\lq micro-springs\rq\rq. 
If a fraction of the wire cross section is worn by corrosion, 
the \lq\lq micro-springs\rq\rq \,in this section do not contribute to the wire elastic constant any more, i.e. the elastic constant of the wire is reduced. 
Hence, for a fixed external stress, wire breaking should occur at a lengthening lower than the nominal breaking stretch. In this scheme the ratio $\lambda/\lambda_{max}$ between the stretch where the breaking effectively occurs $\lambda$ 
and the nominal breaking stretch $\lambda_{max}$ (which for the properties of the MEG\,II wires is close to the elastic limit stretch) is equal to the healthy fraction of the cross section: a strong corrosion is needed to break the wire at small stretches, while a relatively small corrosion is enough to break it when the stretch is not far from the breaking stretch. 
We can write:
\begin{equation}
\label{eq:lambdarat}
\frac{\lambda}{\lambda_{max}} = \frac{\Delta A_{good}}{A} = 1 - \frac{\Delta A_{bad}}{A} \Longrightarrow \Delta A_{bad} = A \left( 1 - \frac{\lambda}{\lambda_{max}} \right)
\end{equation}
where $A$ is the wire cross section and $\Delta A_{bad}$ and $\Delta A_{good}$ are the portions of this cross section 
affected or not by corrosion ($\Delta A_{bad} + \Delta A_{good} = A$). At fixed $\lambda$ wire breaking takes place when, because of corrosion, the portion $\Delta A_{bad}$ reaches the value in formula~\eqref{eq:lambdarat}. 
Corrosion development can be expressed by the amount of material affected as a function of time, which for aluminum 
is well approximated by square root or logarithmic formulas~\cite{corrpaper2}, as:
\begin{equation}
\label{eq:penetration}
\begin{split}
Y \left( t \right) = \sqrt{a + b t} 
\\
Y \left( t \right) = k_{1} \ln{ \left( k_{3} + k_{2} t \right) } 
\end{split}
\end{equation}
where $t$ is the time from the start of corrosion process and $Y$ is the mass
loss per unit surface area measured in ${\rm g} \times \rm{\mu m^{-2}}$. 
Both formulae in~\eqref{eq:penetration} can be used to perform an empirical fit of our data; we chose the square root formula which provides the better interpolation quality, even if satisfactory results are obtained also by using the logarithmic formula.   

Dividing by the aluminum density $\rho_{\rm Al}$ we can compute the depth of corrosion $y$ from the crack point (measured in $\rm{\mu m}$) as:
\begin{equation}
\label{eq:penetration2}
y \left( t \right) = Y\left( t \right)/\rho_{\rm Al} 
\end{equation}
If we assume that corrosion penetrates the material uniformly from the surface towards the core, as if one cuts a triangular piece from a circular loaf of cheese, $\Delta A_{bad} \propto y^{2}$ and one can re-write the equation~\eqref{eq:lambdarat} as:
\begin{equation}
\label{eq:lambdarat2}
\Delta A_{bad} = \gamma y \left( t \right) ^{2} = A \left( 1 - \frac{\lambda}{\lambda_{max}} \right) \Longrightarrow t \left( \lambda \right) = \frac{\frac{A \left( 1 - \frac{\lambda}{\lambda_{max}} \right)} {\gamma} - a}{b}
\end{equation}
where $\gamma$ is the proportionality constant between $\Delta A_{bad}$ and $y^{2}$ and the first formula in~\eqref{eq:penetration} is used as previously stated. 
$t \left( \lambda \right)$ is the time needed for the corrosion process to affect a large enough amount of material and cause the wire to break at the stretch length $\lambda$. 
Therefore, all wires where 
corrosion started and which were exposed to atmospheric relative humidity for a time at least equal to $t \left( \lambda \right)$ are expected to break at stretch $\lambda$ and their number $N \left( \lambda \right)$ can be computed as follows:
\begin{equation}
\label{eq:nlambda}
N \left( \lambda \right) = \alpha \times \sum_{i=1}^{N_{w}} \left( t_{i} - t \left( \lambda \right) \right) \Theta \left( t_{i} - t \left( \lambda \right) \right)
\end{equation}
where $\alpha$ is a proportionality constant, $N_{w}$ is the total number of wires, $t_{i}$ is the relative humidity exposure time of the i-th wire and the Heaviside function $\Theta \left( t_{i} - t \left( \lambda \right) \right)$ ensures that only wires exposed to the relative humidity for a time longer than the minimum are included in the sum. The parameters $\gamma$, $a$, $b$ and $\alpha$ can be extracted by forming and minimising a $\chi^{2}$ based on the comparison between calculated and measured values of $N\left( \lambda \right)$. 

\begin{figure}[htbp]
\centering 
\includegraphics[width=.58\textwidth]{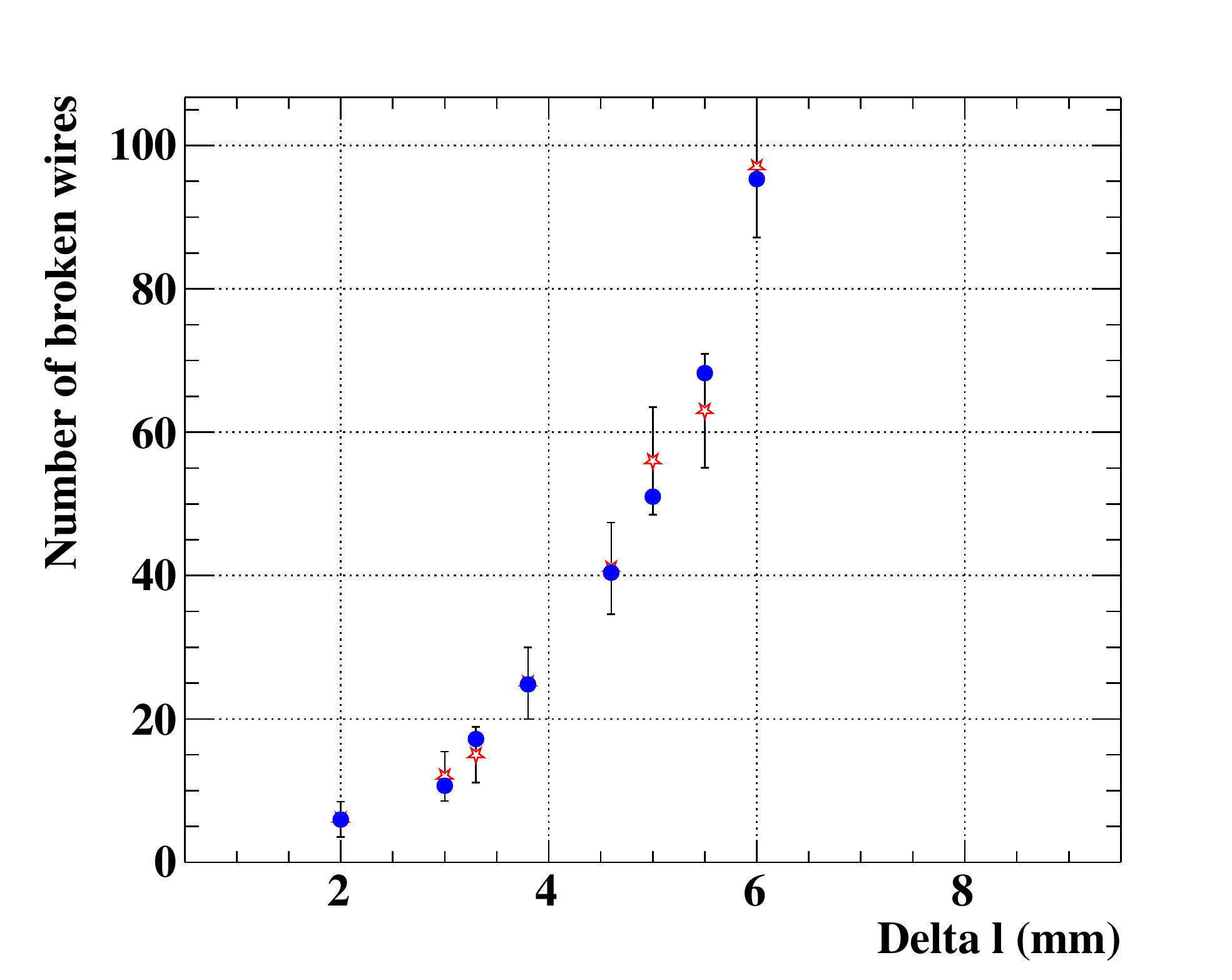}
\caption{\label{fig:model} Integral spectrum of the number of broken wires as a function of stretch. Red open stars are experimental measurements, blue dots are calculated values obtained by $\chi^{2}$ minimisation of formula~\eqref{eq:nlambda}.}
\end{figure}

Figure~\ref{fig:model} shows the comparison between the measured integral distribution and the model results for $N \left( \lambda \right)$ as a function of the stretch: the red stars are the experimental measurements and the blue dots are the calculated values. 
The comparison looks good, with $\chi^{2}/DoF = 1.6/4$, showing that the model is able to reproduce the actual number of broken wires as a function of  stretch. The fit results are reported in table~\ref{tab:nvslambdafit}. 
\begin{table}[htbp]
\centering
\caption{\label{tab:nvslambdafit} Fit parameters of the $N \left( \lambda \right)$ model obtained by $\chi^{2}$ minimisation.}
\smallskip
\begin{tabular}{|c|c|}
\hline
Parameter & Parameter value \\
\hline
$\alpha$    & $\left( 8.4 \pm 0.6 \right) \times 10^{-2}~{\rm days}^{-1}$   \\
$\gamma$ & $\left( 3.0 \pm 0.3 \right)$ \\
$a$           & $\left( 5.4 \pm 0.5 \right) \times 10^{-1}~{\rm \mu m}^{2}$ \\
$b$           & $\left( 9.5 \pm 0.2 \right) \times 10^{-1}~{\rm \mu m}^{2}/{\rm days}$ \\
\hline
\end{tabular}
\end{table}
The model prediction capabilities were tested 
when increasing wires stretch length: for instance in passing from $5$ to $6~{\rm mm}$ the predicted number (previous to the operation) of wires to be removed 
was $38$, against the $41$ observed.  
\section{A different type of wire for a backup drift chamber}
\label{sec:future}
\begin{figure}[htb]
\centering 
\includegraphics[width=.6\textwidth]{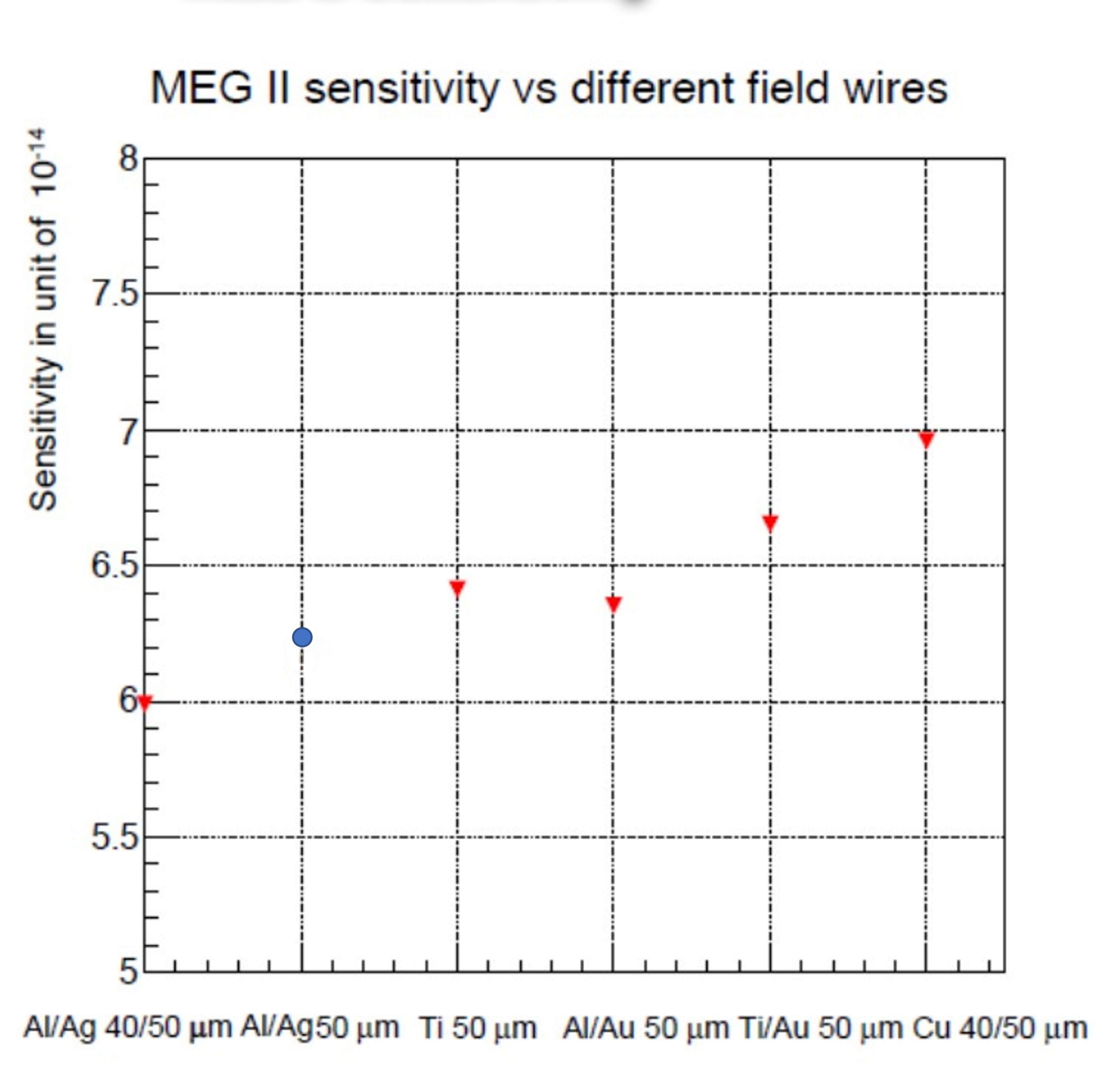}
\caption{\label{fig:sensitivity} The final sensitivity of the experiment versus the use of different kinds of wire. In order: (1) the current wires; (2) Ag coated $50~{\rm{\mu m}}$ Al($5056$) only; (3) uncoated $50~{\rm{\mu m}}$ titanium wire; (4) Au coated $50~{\rm{\mu m}}$ Al wire; (5) Au coated $50~{\rm{\mu m}}$ titanium wire; (6) copper wire with the same diameter of the current ones. The Ag coated $50~{\rm{\mu m}}$ Al($5056$) wire is indicated with a different symbol (a blue dot instead of a red triangle) to stress that it corresponds to the adopted solution.}
\end{figure}

We took advantage of the experience gained with the present chamber to optimise the choice and the installation procedures of wires in the new backup drift chamber, whose construction has already started. The constraints for the construction of a new chamber were far from being irrelevant:
\begin{enumerate}
\item{The analyses and the considerations presented in the previous sections of the paper suggest the necessity of different, more robust wires, possibly insensitive to relative humidity.}
\item{The final goal of the experiment (its sensitivity to $\mu^{+} \rightarrow {\rm e^{+}} \gamma$) should worsen at most by $\approx 10\,\%$ with the use of a new wire.}
\item{The schedule of the experiment implies the construction method of the chamber not change. Therefore soldering the wires to the PCBs is required for anchoring them to the end-plates.}
\end{enumerate}
A Monte Carlo estimate of the experiment sensitivity\footnote{The \lq\lq sensitivity to $\mu^{+} \rightarrow {\rm e^{+}} \gamma$\rq\rq~is defined as the $90\,\%$ Confidence Level upper limit on the branching ratio of the $\mu^{+} \rightarrow {\rm e^{+}} \gamma$ process with respect to the usual muon decay $\mu^{+} \rightarrow {\rm e^{+}} \nu_{\rm e} \bar{\nu}_{\rm \mu}$ under the null signal hypothesis.} as a function of different possible kinds of wire is shown in figure\,\ref{fig:sensitivity}\footnote{This figure was produced by simply changing the wire material in the
simulation: changing wires could however imply a re-design of the mechanics of the drift chamber as in the case of copper wires with a big effect on sensitivity.}.

The use of copper wires clashes with our second criterion; the larger elastic constant of copper wires,
relative to aluminum, would not only imply a larger distortion of the end plates and therefore the need of a more robust material but above all the use of a stronger carbon fiber cover for the drift chamber.
A thicker carbon fiber cover would in turn imply a lower positron reconstruction efficiency, hence a lower sensitivity, not taken into account in the Monte Carlo study used for producing figure\,\ref{fig:sensitivity}.

Titanium wires do exist but there is no producer of gold or silver coated titanium wires. Pure titanium is even more difficult to solder than uncoated aluminum. 

$50~{\rm{\mu m}}$ uncoated Al($5056$) wires are produced
by CFW: we verified that these wires
do not break even after being immersed for several days in pure water. Therefore the possibility of soldering uncoated $50~{\rm{\mu m}}$ Al($5056$) wires was explored in great detail. 
We tested ultrasonic soldering, and used 
all the possible solder pastes currently available on the market which are advertised for soldering pure aluminum or aluminum alloys, in our semi-automatic soldering system.
Unfortunately, given the small diameter of our wires, either soldering is not good enough and the wire slips through the soldering when a mechanical tension above a certain threshold is applied or the wire is effectively soldered, using the strongest pastes containing corrosive fluxants, but, though the soldering spot was carefully washed with isopropyl alcohol, after few days the wire breaks at the soldering due to the ongoing effect of fluxants.
A possible mixed solution with one soldering point for electrical contact and glueing to a groove on the PCB for mechanical tension was found but would
have required exceedingly long times to be fully tested, which would have delayed too much the new chamber construction.

\begin{figure}[htb]
\centering 
\includegraphics[width=.6\textwidth]{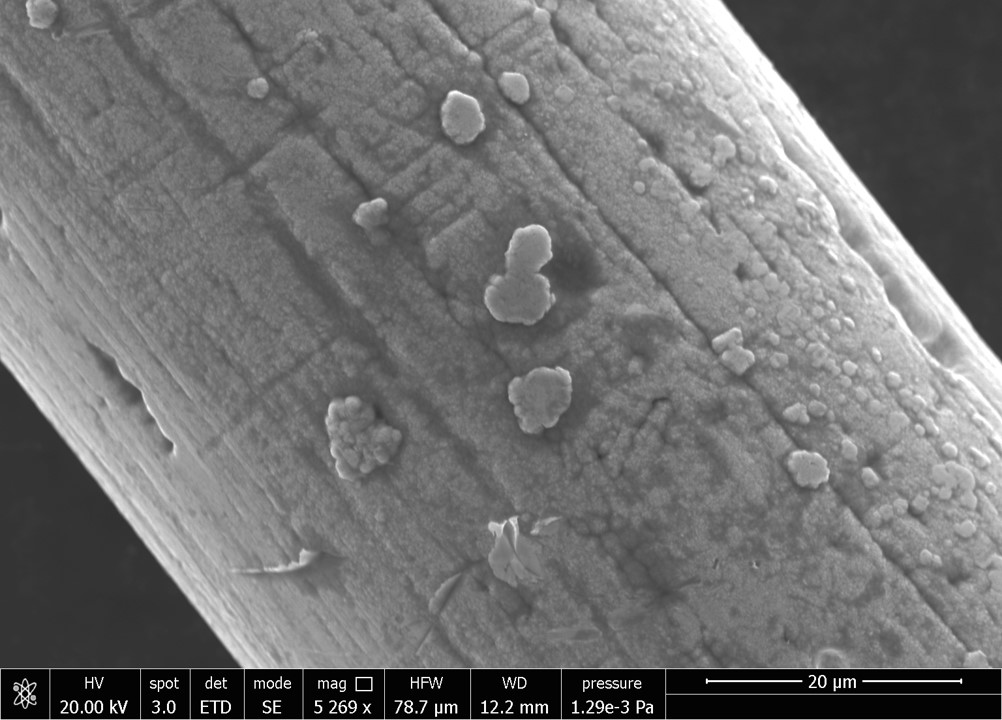}
\caption{\label{fig:nonultra} SEM picture of a non ultra-finished $50~{\rm{\mu m}}$ Ag-coated Al($5056$) wire.}
\end{figure}

As already noted above the scrutiny of our Ag-coated aluminum wires under an SEM shows that the density of cracks in the coating varies very much among the different spools of wire received. This was also tested by immersing samples from different spools in pure water for several hours. 
In some cases the sample was completely destroyed while in others it just showed a limited number of points at which gas bubbles, hydrogen presumably, formed.
After discussion with the producer we presume that these differences are due to the last
step of production (ultra-finishing) which is not very well controlled.
A sample of $50~{\rm{\mu m}}$ Ag-coated aluminum wire produced without the final ultra-finishing step does not show visible cracks at SEM (see figure \ref{fig:nonultra}) though its surface is a little less regular than the ultra-finished samples and when immersed in pure water it shows just a few points affected by water. 
The different sensitivity to relative humidity of the different wires spools employed in the present chamber
can also explain why the relative humidity dependence of broken wires shown in figure\,\ref{fig:humid} is
not so regular: a general linear dependence can be seen but it is affected by the different sensitivity to relative humidity of the spools used for the particular group of layers.

For the new chamber construction it was decided to proceed with the use of non ultrafinished $50~{\rm{\mu m}}$ Ag-coated aluminum wires for all cathodes and to protect the wires by using dry air (low relative humidity) as much as possible during the assembly phase. 
The sensitivity expected with this kind of wire is marked in figure\,\ref{fig:sensitivity} with a 
blue dot and fits well with our requirements. By planning the construction appropriately we estimate the time of exposure to relative humidity to be between a few days for the outermost wire layers to at most two months for the innermost ones.

An upper limit on the expected number of broken wires in the new chamber can be obtained by using the model described in section~\ref{sec:ana}, the estimated time of exposure to humid atmosphere and the relative breaking probability $P_{50/40}$ computed in formula~\eqref{eq:relprob}. The result is that no more than $2 - 3$ wires should break in the new chamber at the operating stretch of $5.2~{\rm mm}$. This number is a very conservative estimate, since the effect of avoiding the ultra-finishing procedure is not taken into account. By exploiting the strong dependence of the number of breakages on the stretch the few possible \lq\lq weak\rq\rq~wires could be forced to break by an appropriate overstretch immediately after the chamber assembly phase; after removing the possible few broken wires the chamber will be sealed and flushed with a dry gas mixture for normal operation, preventing any further breakages.
\section{Summary and conclusions.}
\label{sec:summary}
Several Ag-coated aluminum cathode wires, with 
$40$ and $50~{\rm{\mu m}}$ diameters, of the MEG\,II drift chamber broke during the assembly phase of the chamber. 
The wire fragments could be extracted without any further damage to the chamber. Detailed analysis of the breakages showed that the cause was exposure of wires to relative humidity during construction of the chamber. The main cause of the strong wire sensitivity to relative humidity is the final production step named ultra-finishing. A phenomenological model of the
number of breakages as a function of exposure to relative humidity and mechanical tension was developed which matches very well the experimental observations. Construction of a new backup drift chamber has started; the new cathodes which will be used are
$50~{\rm{\mu m}}$ Ag-coated aluminum wires by the same company in which the ultra-finishing step is avoided.
The empirical model developed for predicting the breakages-relative humidity-stress relationship was used to set up 
an assembly operational plan for the new chamber which foresees a limited exposure to relative humidity of the wires. 

\acknowledgments{We are very grateful to G.Balestri, A.Bianucci, M.D'Elia, A.Innocente, A.Miccoli, C.Pinto, G.Petragnani and A.Tazzioli for their invaluable work throughout all the construction period of our drift chamber which lasted for a considerable time.}

\end{document}